\documentclass{article}
\usepackage{graphicx}
\usepackage{color}

\def\m{{\rm m}}

\def\de{{\rm de}}
\newcommand{\ben}{\begin{eqnarray}}
\newcommand{\een}{\end{eqnarray}}
\newcommand{\be}{\begin{equation}}
\newcommand{\ee}{\end{equation}}
\newcommand{\n}{\label}

\newcommand{\ga}{\gamma}

\begin{document}

\title{Bulk viscous cosmological model with interacting dark fluids}

\author{Gilberto M. Kremer\footnote{kremer@fisica.ufpr.br},
       Oct\'avio A. S. Sobreiro\footnote{oass05@fisica.ufpr.br}
\\
               Departamento de F\'{\i}sica,
Universidade Federal do Paran\'a,
\\Caixa Postal 19044, 81531-990
Curitiba, Brazil }

%\date{Received: date / Accepted: date}
% The correct dates will be entered by the editor
\date{}
\maketitle

\begin{abstract}
The objective of the present work is to study a
cosmological model for a spatially flat Universe whose constituents are a dark energy field and a matter field which includes baryons and dark matter.
The con\-stitu\-ents are supposed to be in interaction and irreversible processes are taken into account through the inclusion of a non-equilibrium pressure.
 The non-equilibrium pressure is considered to be proportional to the Hubble parameter within the framework of
a first order thermodynamic theory.  The dark energy and matter fields are  coupled by
their barotropic indexes, which are considered as functions of the
ratio between their energy densities. The free parameters of the model are adjusted from the best fits of the Hubble parameter data.
A comparison of the viscous model with the non-viscous one is performed. It is shown that  the equality of the dark energy and  matter density parameters
 and the decelerated-accelerated transition  occur at earlier times when the irreversible processes are present. Furthermore,  the density
and deceleration parameters and the  distance modulus have the correct
behavior which is expected for a viable  scenario of the present status of the Universe.


\end{abstract}

%%%%%%%%%%%%%%%%%%%%%%%%%%%%%%%%%%%%%%
\section{Introduction}
%%%%%%%%%%%%%%%%%%%%%%%%%%%%%%%%%%%%%%
\label{intro}

The cosmic observations from the type Ia  supernovae \cite{A} suggest that the present period of the Universe is experimenting an  accelerated expansion. Since  matter contributes with attractive forces and positive pressure which decelerate the expansion, an exotic component -- the so-called dark energy -- with negative pressure must be postulated to take into account the present accelerated expansion. Another dark component is also necessary to explain the measurements of rotation curves of spiral galaxies \cite{B}. This component, called dark matter, interacts only gravitationally with ordinary matter. Dark energy can be modeled by a cosmological constant \cite{C}, however, it suffers the so-called fine-tuning and cosmic coincidence problems \cite{D}. Hence, several models for the dark energy having dynamical properties were analyzed in the literature, among others we cite: scalar fields, tachyon fields, fermion fields, phantom fields, exotic equations of state  and so on.

 It is expected that the dark components do not evolve separately. Indeed it is known that the problems stated above have a promising resolution if we take into account a dark energy-dark matter interaction. This interaction is supposed to be negligible at high red-shifts while it is preponderant at lower red-shifts. This may also alleviate the coincidence problem in the sense that it is possible to choose an appropriate form for the interaction term leading to a nearly constant ratio   between the energy density of the matter field and the one of the dark energy at low red-shifts. Several cosmological models were proposed with interacting dark components, among others we quote the works given in the reference \cite{E}. On the other hand, it is also understood that irreversible processes in the evolution of the Universe may also contribute significantly for the alleviation of the coincidence problem. In the Friedmann-Robertson-Walker metric this is effectively done through the introduction of a bulk viscosity associated to a non-equilibrium pressure. (see e.g. \cite{F}).

The aim of this work is to develop a cosmological model for a spatially flat Universe with interacting dark components where irreversible processes are considered.
We follow \cite{CFK} and couple the dark energy and matter fields  by
their barotropic indexes, which are considered as functions of the
ratio between their energy densities. This is in contrast with most works in the literature, which directly consider an explicit form for the interaction term.  Furthermore, we introduce a non-equilibrium pressure -- within the framework of a first order thermodynamic theory -- which is the responsible for the irreversible processes.

The work is structured as follows.  In section \ref{diss} the general features of the proposed model for the Universe -- where dissipative effects are present and with interacting dark fluids -- is discussed. The analysis of  the cosmological constraints and  cosmological solutions which follow from the  model is the subject of section \ref{cos}.  Finally, in section \ref{conc} we present our conclusions.

%%%%%%%%%%%%%%%%%%%%%%%%%%%%%%%%%%%
\section{Dissipative interacting dark fluids}
%%%%%%%%%%%%%%%%%%%%%%%%%%%%%%%%%%%
\label{diss}

Let us consider a homogeneous, isotropic and spatially flat Universe described by the  Friedmann-Robertson-Walker metric
$ds^2=dt^2-a(t)^2(dx^2+dy^2+dz^2)$ where $a(t)$ denotes the cosmic scale factor. Furthermore, let us consider a cosmological model where the Universe is modeled as  a mixture of two constituents, namely, dark energy (de) and matter (m) which represents the baryons and the dark matter.  In this model irreversible processes are taken into account by considering  a non-equilibrium pressure and it is supposed to exist an energy transfer between the dark energy and the matter field. The Friedmann equation and the evolution equation for the total energy density $\rho=\rho_\m+\rho_\de$ read
 \ben\n{1}
 3H^{2}=\rho_\m+\rho_\de,\\\n{1a}
\dot\rho_\m+\dot\rho_\de+3H(\rho_\m+\rho_\de+p_\m+p_\de+ \varpi)=0.
 \een
 In the above equations $H=\dot a/a$ denotes the Hubble parameter, $p=p_\m+ p_\de$ is the total equilibrium pressure and $\varpi$ stands for the non-equilibrium pressure, also known as the dynamic pressure.

 We follow \cite{CFK} and decouple (\ref{1a}) into two ``effective conservation equations'', namely,
 \ben\n{2}
 \dot\rho_\m+3H\ga_\m^e\rho_\m=0,\qquad
 \dot\rho_\de+3H\ga_\de^e\rho_\de=0.
 \een
 Above, it was introduced the effective barotropic indexes $\ga_i^e \,(i=\m, \de)$ related by
 \ben\n{3}
 \ga_\m^e&=&\ga_\m+\frac{\ga_\de-\ga_\de^e}{r} +\frac{\varpi}{\rho_\m},
 \een
 where $r=\rho_\m/\rho_\de$ denotes the ratio between the energy densities and
 $\gamma_i \, (i=\m, \de)$ represent  constant barotropic indexes of the equations of state
 $p_i=(\gamma_i-1)\rho_i$. This decoupling is motivated from the fact that we do not assume an explicit form for the interaction term between dark matter and dark energy. Rather we consider this interaction is intrinsically connected to their barotropic indexes.

 Again by following \cite{CFK} we assume that the effective barotropic index of the dark energy is given by
 \be\n{4b}
 \ga_\de^e=\ga_\de-F(r),
 \ee
 where $F(r)$ is a function which depends only  on the ratio of the energy densities $r$. The physical motivation for this choice is given by the interaction between the dark fluids. Indeed while $\gamma_{m}^{e}$ and $\gamma_{de}^{e}$ give the influence of the interaction term in the field equations, $F(r)$ accounts for the nature of this interaction. Since we are concerned with the coincidence problem, it is reasonable to suppose that $F$ depends on the ratio $r=\rho_{m}/\rho_{de}.$ By taking into account the previous representation for $\ga_\de^e$
 we can rewrite (\ref{2})  as
 \ben\n{4c1}
 \dot\rho_\m+3H\ga_\m\rho_\m&=&-3H\rho_\de F-3H\varpi,\\\n{4c2}
 \dot\rho_\de+3H\ga_\de\rho_\de&=&3H\rho_\de F.
 \een

Within the framework of ordinary (first order or Eckart) thermodynamic theory the non-equilibrium pressure (see e.g. \cite{F}) is proportional to the  Hubble parameter $H$ with proportionality factor identified with the coefficient of bulk viscosity $\eta$, i.e., $\varpi=-3\eta H$.  According to kinetic theory of relativistic gases (see e.g. \cite{CK}) the bulk viscosity is proportional to the temperature with an exponent that depends on the intermolecular forces, so that it is usual in cosmology to assume that $\eta\propto\rho^m$, where $m$ is a positive constant.

If we suppose that the coefficient of bulk viscosity is proportional to the square root of the total energy density --  $\eta=\eta_0 \sqrt{\rho}$   with $\eta_0$ a constant --  the field equations are integrable and the expression for the effective barotropic indexes (\ref{3})  become
\ben\n{4}
 \ga_\m^e=\ga_\m+\frac{\ga_\de-\ga_\de^e}{r} -\sqrt{3}\left(1+{1\over r}\right)\eta_0.
 \een
We may infer from (\ref{4b}) and (\ref{4}) that the effective baro\-tropic indexes are functions only of the ratio between the energy densities.

 Now let us analyze the  evolution equation for the  ratio between the energy densities, which is given by
 \ben\n{4c}
 \dot r=-3Hr\mathcal{F}(r),
 \een
 where $ \mathcal{F}(r)$ denotes the expression
 \ben\n{4d}
  \mathcal{F}(r)=\left[\ga_\m-\ga_\de+\left(1+{1\over r}\right)\left(F(r)-\sqrt{3}\eta_0\right)\right].
 \een
 If we assume that a stationary state of the Universe is attained by a constant value of $r=r_s$, this implies that  $ \mathcal{F}(r_s)=0$. Hence,  the constant solutions $r_s$ will be  stable if
 \be\n{in}
\left({d\mathcal{F}(r)\over dr}\right)_{r=rs}\geq0,
\ee
so that we obtain from (\ref{4d})  the inequality
\be\n{5}
r_s\left(1+r_s\right)\left({d {F}(r)\over dr}\right)_{r=rs}-\left(F(r_s)-\sqrt{3}\eta_0\right)\geq0,
\ee
by taking into account that the barotropic indexes $\ga_\m$ and $\ga_\de$ are constants. From the inspection of  (\ref{5})  we may infer that the simplest choice  $F=\sqrt{3}\eta_0$ fulfills the above inequality. This choice fulfills the stability condition. Moreover, the interaction term in the form $3H\lambda\rho_\de$, with $\lambda$ a  constant and proportional to $\rho_\de$, is consistent with the Le Ch\^{a}telier-Braun principle of thermodynamics as it was shown by \cite{Pavon}.

In order to determine the solutions of the field equations,
we start by analyzing the evolution equation for the energy density of the dark energy (\ref{2})$_2$. According to the ansatz (\ref{4b}) and of the choice of  $F$, the effective barotropic index $\ga^e_\de=\ga_\de-\sqrt{3}\eta_0$ is a constant. Hence, we may integrate equation (\ref{2})$_2$ and obtain
\be\n{2b}
\rho_\de=\rho_\de^0\left({a_0\over a}\right)^{3\ga_\de^e},
\ee
where the index $0$ stands for the present values of the variables.

 From the differentiation of the Friedmann equation (\ref{1}) with respect to time it follows
 \be\label{eq1}
 \dot H+{3\over2}(\gamma_\m-\sqrt{3}\eta_0)H^2-{1\over2}(\gamma_\m-\gamma_\de)\rho_\de^0\left({a_0\over
a}\right)^{3\gamma_\de^e}=0.
\ee
The integration of  the above equation leads to
\ben\label{eq2}
H^2=\mathcal{C}\left({a_0\over
a}\right)^{3(\gamma_\m-\sqrt{3}\eta_0)}+{\rho_\de^0\over3}\left({a_0\over
a}\right)^{3\gamma_\de^e}.
\een
The constant of integration $\mathcal{C}$  is found by considering the current values  of the cosmic scale factor $a_0$ and of the Hubble constant $H_0$, yielding
 \be
\mathcal{C}=H^2_0-{\rho_\de^0\over 3}.
\ee
Hence, (\ref{eq2}) can be rewritten in terms of the red-shift  $z=\left(a_0/a-1\right)$ as
 \be\n{6}
 {H^2\over H_0^2}=\Omega_\m^0\left(1+z\right)^{3(\gamma_\m-\sqrt{3}\eta_0)}
 +\Omega_\de^0\left(1+z\right)^{3\ga_\de^e},
\ee
where $\Omega_i=\rho_i/(\rho_\m+\rho_\de)$ denote the density parameters.

From the knowledge of  $H^2=\left(\rho_\m+\rho_\de\right)/3$, we can obtain the density parameters of the matter and dark energy in terms of the red-shift, namely,
\ben\n{7a}
\Omega_\m (z)={\Omega_\m^0(1+z)^{3(\ga_\m-\sqrt{3}\eta_0)}\over \Omega_\m^0(1+z)^{3(\ga_\m-\sqrt{3}\eta_0)}+\Omega_\de^0(1+z)^{3\ga_\de^e}},\\\n{7b}
\Omega_\de (z)={\Omega_\de^0(1+z)^{3\ga_\de^e}\over
\Omega_\m^0(1+z)^{3(\ga_\m-\sqrt{3}\eta_0)}+\Omega_\de^0(1+z)^{3\ga_\de^e}}.
\een

The determination of the the ratio between the energy densities as function of the red-shift follows from  $r(z)= \Omega_\m (z)/\Omega_\de (z)$.
Furthermore,  since the non-equilibrium pressure is given by $\varpi=-3\sqrt{3}\eta_0H^2$, it can be also expressed as a function of the red-shift thanks to (\ref{6}).

 Another  parameter which is important in cosmology is the deceleration parameter $q=1/2+3w_e/2$,  which is given in terms of the  effective parameter $w_e={(p_\m+p_\de+\varpi)/(\rho_\m+\rho_\de)}$.  From the barotropic equations of state and from the representation of the non-equilibrium pressure  the effective parameter becomes
\be\n{9}
w_e=(\ga_\m-1)\Omega_\m(z)
+(\ga_\de^e+\sqrt{3}\eta_0-1)\Omega_\de(z)-\sqrt{3}\eta_0.
\ee

By inspecting the expressions (\ref{6}) -- (\ref{9}) we may infer that there exist three free parameters in the proposed model, which are the coefficients  $\ga_m$, $\eta_{0}$ and $\ga_\de^e$. In  the next section an analysis to set cosmological constraints on the free parameters is performed and  the cosmological solutions are analyzed.

%%%%%%%%%%%%%%%%%%%%%%%%%%%%%%%%%%%
\section{Cosmological constraints and cosmological solutions}
%%%%%%%%%%%%%%%%%%%%%%%%%%%%%%%%%%%
\label{cos}

The coefficients $\ga_\m$, $\eta_{0}$ and $\ga_\de^e$ can be found from  the observational cosmological constraints which are based on the data of the Hubble parameter $H(z)$ given in Table \ref{tab:1} -- taken from \cite{Ma} -- together with the values $H_0=72$ km/(s Mpc),  $\Omega_\m^0=0.30$ and $\Omega_\de^0=0.70$  \cite{Freedman:2000cf}. The  set of values given in Table \ref{tab:1} was used in the work \cite{CFK} and the adopted methodology is explained in the appendix.

\begin{figure*}
\vskip1cm
\includegraphics[width=0.4\textwidth]{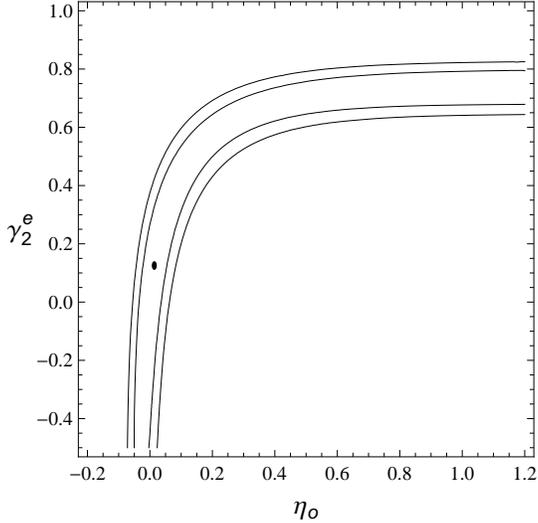}
\caption{Confidence regions for the best fit values for the viscous case.}
\label{fig:1}
\vskip1cm
\end{figure*}

For the viscous case, we have considered a dust-like matter field ($\gamma_{m}=1$) and adjusted the parameters $\gamma_\de^e$ and $\eta_0$. In Figure \ref{fig:1} it is plotted the probability ellipsis in the plane $\gamma_\de^e$ versus  $\eta_0$ and the best fit value is indicated by a dot, which corresponds to $\gamma_\de^e=0.125445$ and $\eta_0=0.0140124$ with
$\chi^2=9.104007$. 

\begin{figure*}
\vskip1cm 
\includegraphics[width=0.4\textwidth]{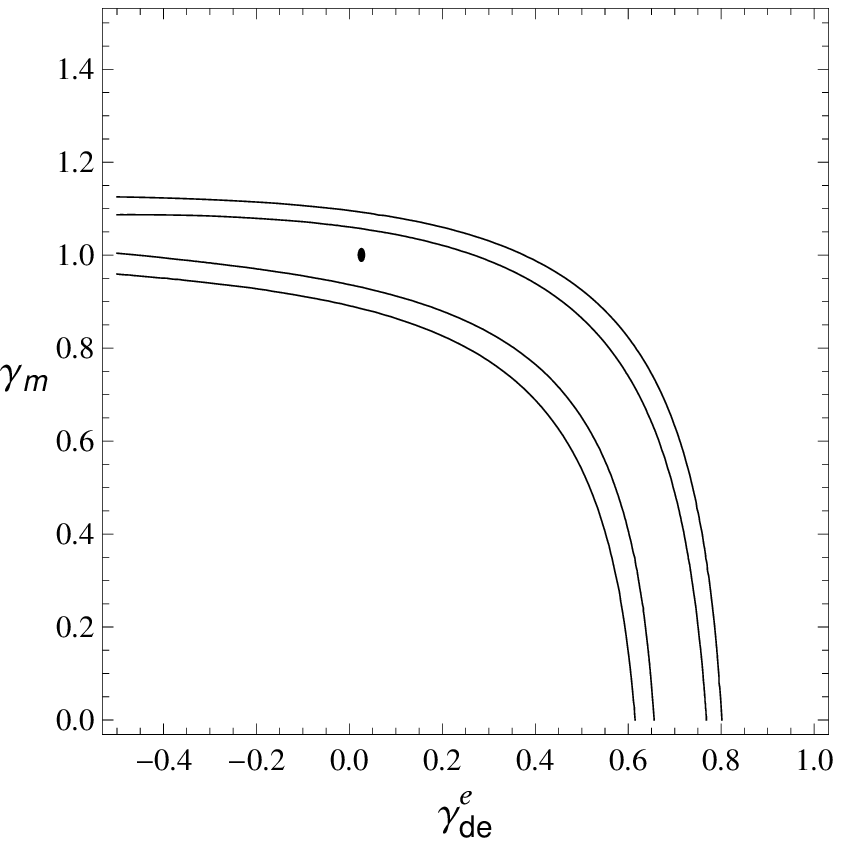}
\caption{Confidence regions for the best fit values for the non-viscous case.}
\label{fig:2}
\vskip1cm
\end{figure*}

In order to interpret the results for the viscous case, we compare it with the non-viscous one, which refers also to a non-interacting model. In this case the free parameters are $\gamma_\m$ and $\gamma_\de^e$, and in Figure \ref{fig:2} we show the  probability ellipsis in the plane $\gamma_\de^e$ versus  $\gamma_\m$. The best fit value is indicated by a dot, which corresponds to $\gamma_\de^e=0.0259052$ versus  $\gamma_\m=1.0051$ with $\chi^2=9.1407510$.

In Figures \ref{fig:1} and \ref{fig:2} the points inside the inner ellipses or between them stand for the true values of parameters with $68.3\%$ and $95.4\%$ which correspond to $1\sigma$ and $2\sigma$ confidence regions, respectively.

\begin{table}
\caption{Hubble parameter $H(z)$ from \cite{Ma}.}
\label{tab:1}       % Give a unique label
\begin{tabular}{lll}
\hline\noalign{\smallskip}
$z$ & $H(z)$ & $1\sigma$  \\
 & km/(Mpc $\,$s)&uncertainty\\
\noalign{\smallskip}\hline\noalign{\smallskip}
0.09& 69 & $\pm12$ \\
0.17&83&$\pm$8.3\\
0.27&70&$\pm$14\\
0.40&87&$\pm$17.4\\
0.88& 117&$\pm$23.4\\
1.30&168&$\pm$13\\
1.43&177&$\pm$14.2\\
1.53&140&$\pm$14\\
1.75&202&$\pm$40.4\\
\noalign{\smallskip}\hline
\end{tabular}
\end{table}

\begin{figure*}
\vskip1cm
\includegraphics[width=0.4\textwidth]{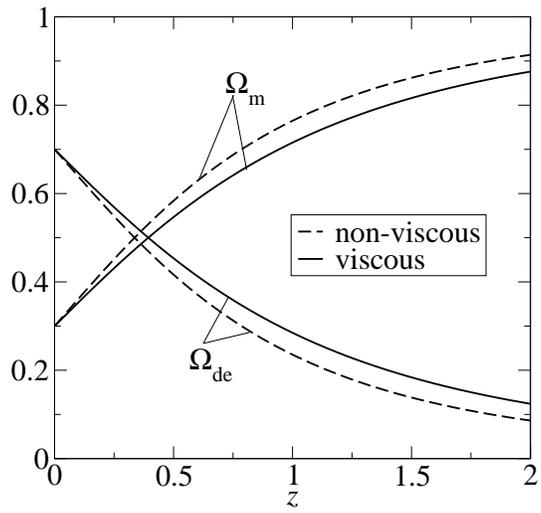}
\caption{Density parameters as functions of the red-shift $z$. Solid lines - viscous; dashed lines - non-viscous.}
\label{fig:3}
\vskip1cm
\end{figure*}

\begin{figure*}
\vskip1cm
\includegraphics[width=0.4\textwidth]{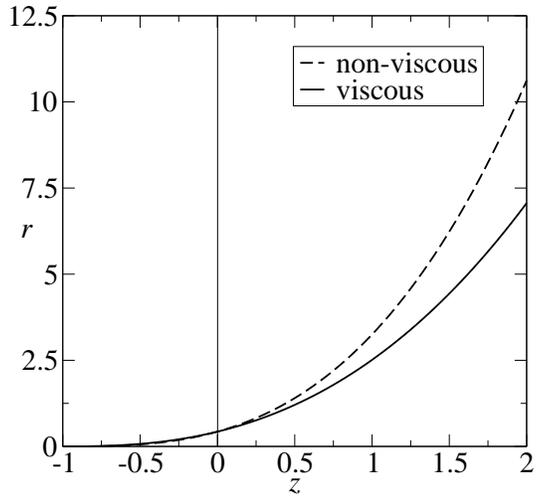}
\caption{Ratio between dark matter and dark energy as function of the
red-shift $z$. Solid lines - viscous; dashed lines - non-viscous.}
\label{fig:4}
\vskip1cm
\end{figure*}

From the knowledge of free parameters
 of the model it is  possible to perform an analysis of the cosmological solutions. 
We start with the investigation of the density parameters which are plotted
as functions of the red-shift in Figure \ref{fig:3}, the solid
lines corresponding to the viscous case, whereas the dashed lines to
the non-viscous case. One can infer from this figure that the energy
transfer from the dark energy to the matter field is more pronounced for
the non-viscous case, since for this case the growth of the density
parameter of the matter field and the corresponding decay of the dark
energy with the red-shift are more pronounced than those for the
viscous case. This behavior is expected when we analyze the evolution equations for the energy densities (\ref{4c1}) and (\ref{4c2}) for the viscous case and 
can also be verified from the analysis of Figure \ref{fig:4}
which represents the evolution of the ratio of the two energy
densities $r = \rho_\m/\rho_\de$ with the red-shift. This last figure
also shows that in the future -- i.e., for negative values of the
red-shift -- there is no difference between the two cases, since both
tend to small values, indicating a predominance of the dark energy
in the future.

\begin{figure*}
\vskip1cm
\includegraphics[width=0.4\textwidth]{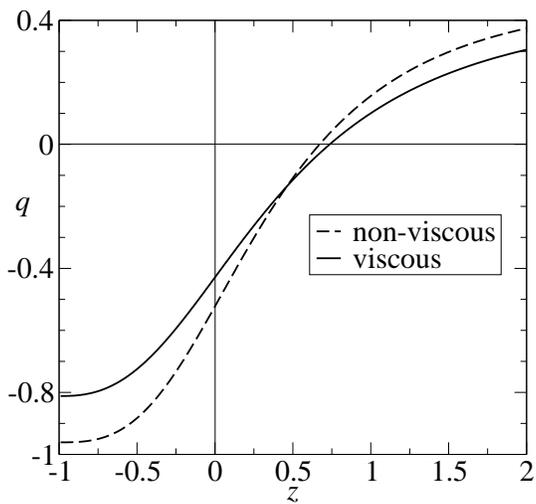}
\caption{Deceleration parameter as function of the
red-shift $z$. Solid
lines - viscous; dashed lines - non-viscous.}
\label{fig:5}
\vskip1cm
\end{figure*}

In Figure \ref{fig:5} the deceleration parameter  is plotted
for the two cases. The present values of the deceleration parameter
$q(0)$ and the value for the red-shift $z_t$ where the transition
from a decelerated to an accelerated regime occur are : (i)
$q(0)\approx-0.43$ and $z_t\approx0.74$ for the viscous case
and (ii) $q(0)\approx-0.52$ and $z_t\approx0.67$ for the
non-viscous case. These values are of the same order of magnitude of
the  values given in the literature: $q(0)=-0.46\pm0.13$ (see~\cite{Vir}) and
$z_t=0.74\pm0.18$ (see~\cite{riess}). When the viscous case is compared with the non-viscous one we may infer from this figure  that: (i) the former has a smaller deceleration in the past than the latter; (ii) the transition from a decelerated to an accelerated regime occurs earlier for the former and (iii) in the present and in the future the former has a smaller acceleration than the latter.

\begin{figure*}
\vskip1cm
\includegraphics[width=0.4\textwidth]{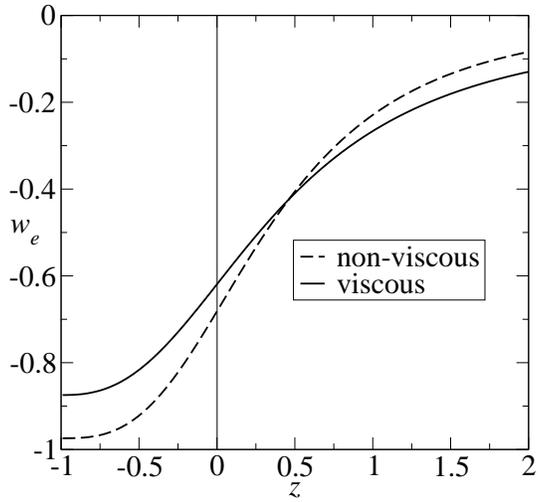}
\caption{Effective index $w_e$ as function of the
red-shift $z$. Solid lines -
viscous; dashed lines - non-viscous.}
\label{fig:6}
\vskip1cm
\end{figure*}

The effective index $w_e$ as function of the red-shift is shown in Figure \ref{fig:6} for the viscous and non-viscous cases. We may conclude that in the future the mixture of matter and dark energy behaves  like a quintessence and the non-viscous case approximates to a cosmological constant with $w_e\approx-1$. For large values of the red-shift $w_e$ tends to zero for the non-viscous case and to a small negative value for the viscous one. Here we call attention that for large values of the red-shift it is necessary to include a radiation field which will imply  in a positive effective index.

\begin{figure*}
\vskip1cm
\includegraphics[width=0.4\textwidth]{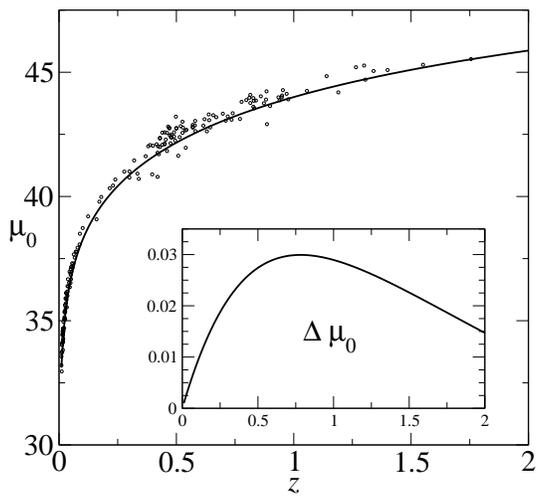}
\caption{Distance modulus $\mu_0$ as function of the red-shift $z$. Small frame:  $\Delta\mu_0=\mu_0^{\rm viscous}-\mu_0^{\rm non-viscous}$ .}
\label{fig:7}
\vskip1cm
\end{figure*}

In Figure \ref{fig:7} it is represented the  distance modulus $\mu_0$, which is the
difference between the apparent magnitude $m$ and the absolute
magnitude $M$ of a source. Its expression is given by
\be
 \mu_0=m-M=5\log \left\{(1+z)\int_0^z{dz'\over
 H(z')}\right\}+25,
 \ee
 where the quantity within the braces represents the luminosity distance in  Mpc. The
 circles in this figure  are  observational data for super-novae of type Ia taken from the
 work \cite{14}. This reference contains 4 different data sets related to various light-curve fitters. For practical purposes, we adopted the SALT data set ($R_{V}=3.1$). It is possible to conclude that there is a good fitting of the curve with the observational data. Moreover, it can be seen from the small frame in this figure that there is no sensible difference between the curves for the viscous and non-viscous cases.

%%%%%%%%%%%%%%%%%%%%%%%%%%%%%%%%%%%
\section{Conclusions}
%%%%%%%%%%%%%%%%%%%%%%%%%%%%%%%%%%%
\label{conc}

In this work we studied a cosmological model with
interacting dark fluids in a dissipative Universe where  the
non-equilibrium pressure is the responsible for the irreversible processes. The non-equilibrium pressure was supposed to be proportional to the Hubble parameter within the framework of
a first order thermodynamic theory. The coupling between matter and dark energy was
made through their barotropic indexes, which were considered as
functions of the ratio between their energy densities. The function of the ratio between the energy densities -- which is the responsible for the energy transfer between matter and dark energy -- follows from the stability analysis of the differential equation for the density ratio. A procedure  was performed to set observational constraints on the
free parameters of the model by using the observational data of the Hubble parameter. It was shown that the energy transfer from the dark
energy to the matter field is more efficient for the non-viscous
case. Furthermore, for both the viscous and non-viscous cases we obtained that the dark energy density predominates
in the future,  the mixture behaves like a quintessence in the future and the  values of the
deceleration parameter are of the same order as those given in the literature. It was shown also that the behavior of
the distance modulus $\mu_0$ -- which is related with the luminosity distance -- has
a good fit with the observational  values.

\section*{Acknowledgements}
GMK acknowledges the fruitful discussions with Luis P. Chimento and M\'onica Forte and the support by CNPq. OASS acknowledges the support by CAPES.

%%%%%%%%%%%%%%%%%%%%%%%%%%%%%%%%%%%
\section*{Appendix: Bayesian Inference}
%%%%%%%%%%%%%%%%%%%%%%%%%%%%%%%%%%%
In a statistical sense a physical model may be thought as described by a set of parameters. The determination of these parameters may be carried out in many ways; the most commonly used framework to accomplish this is Bayesian inference, a well-known method of statistical inference which employs evidence to estimate parameters of a model. The main purpose of this section is just to give a brief introduction to the subject.

For a given model and data set, Bayesian inference employs a probability distribution called \emph{posterior} probability to summarize all uncertainty. This probability distribution is proportional to a prior probability distribution (or simply the prior) and a likelihood function. The later, denoted by $\mathcal{P}(\mathbf{D}|\mathbf{\theta})$, is usually defined as the unnormalized probability density of measuring the data $\mathbf{D}=\{D_{1},D_{2},...,D_{n}\}$ for a given model $\mathcal{M}$ in terms of its parameters $\bf{\theta}=\{\theta_{1},\theta_{2},...,\theta_{n}\}$. For our purposes it suffices to assume that the measured values are normally distributed around their true value, so that \be
\mathcal{P}(\mathbf{D}|\mathbf{\theta}) \propto \exp \left[-\chi^{2}(\mathbf{\theta})/2
\right]. \ee

The posterior $\mathcal{P}(\theta|\mathbf{D})$ is determined by Bayes' theorem \be
\mathcal{P}(\theta|\mathbf{D})=\frac{\mathcal{P}(\mathbf{D}|\theta)\mathcal{P}(\theta)}{\int
d\theta \mathcal{P}(\mathbf{D}|\theta)\mathcal{P}(\theta)}, \ee where $\mathcal{P}(\theta)$ denotes the prior probability distribution. The prior carries all previous knowledge about the parameters before the measurements have been performed.

Parameter estimation is performed in Bayesian inference by maximizing the posterior $\mathcal{P}(\theta|\mathbf{D})$. This is in contrast with the frequentist approach, in which the likelihood $\mathcal{P}(\mathbf{D}|\mathbf{\theta})$ is maximized. Nevertheless, whenever  the so-called uninformative priors are considered, both frameworks lead to the same conclusions. If the measured data are independent from each other as well as Gaussian distributed around their true value, $\mathbf{D}(\theta)$, then maximizing the likelihood $\mathcal{P}(\mathbf{D}|\mathbf{\theta})$ is equivalent to minimize the chi-square function \be \chi^{2}(\theta)
\equiv (\mathbf{D}^{obs}-\mathbf{D}(\theta))C^{-1}(\mathbf{D}^{obs}-\mathbf{D}(\theta))^{T}, \ee where $C$ is the covariance matrix given by the experimental errors. For uncorrelated data $C_{ij}=\delta_{ij}\sigma^{2}_{i}$ and \be \chi^{2}(\theta) \equiv
\sum_{i=1}^{n}\left(\frac{D^{obs}-D(\theta)}{\sigma^{2}_{i}}\right)^{2},
\ee where $\sigma_{i}$ denotes the experimental errors.

In Bayesian inference, the confidence intervals are drawn around the maximal likelihood point, giving the best fit parameters. It is conventionally used $1\sigma$ and $2\sigma$ confidence regions with $68,3\%$ and $95,4\%$ of probability, respectively, for the true value of parameters. These regions are mathematically defined by the inequalities \be \chi^{2}(\theta)-\chi^{2}(\theta_{bf}) \leq 2.3, \ee for $1\sigma$ range and \be \chi^{2}(\theta)-\chi^{2}(\theta_{bf}) \leq 6.17, \ee for $2\sigma$ range, where $\theta_{bf}$ denotes the best fit value of parameters.

\end{document}